# First-principles calculations of $Sc_2CdS_4$ and $Y_2CdS_4$ compounds

Abdul Ahad Khan[1], Zubaida Noor[2], G. Murtaza[2,3,]

[1]*Department of Physics, University of the Peshawar, Khyber Pakhtunkhwa, Pakistan*

[2]*Materials Modeling Lab, Department of Physics, Islamia College Peshawar 25120, Khyber Pakhtunkhwa, Pakistan*

[3]*Department of Mathematics & Natural sciences, Prince Mohammad Bin Fahd University, P. O. Box 1664, Alkhobar 31952, Kingdom of Saudi Arabia*

**Abstract**

Direct energy bandgap materials are crucial for the efficient optoelectronics devices. Therefore, the investigation of new direct gap materials is important. In the present work, two novel *d*-metal sulfides $Sc_2CdS_4$ and $Y_2CdS_4$ compounds are investigated by using the all electron full potential linearized augment plane wave method. Both the compounds show semiconducting nature and direct band gap with a value 1.886eV for $Sc_2CdS_4$ and 2.209eV for $Y_2CdS_4$, respectively. Strong hybridization between S-*p* and Sc/Y-*d* orbitals present among valence and conduction bands which is beneicial to electrical transport. Key optical parameters are calculated. The static value of the reflectivity *R*(0) and refractive index *n*(0) are vary inversely with the energy band gap ($E_g$). Both the compounds $Sc_2CdS_4$ and $Y_2CdS_4$ are P-type thermoelectric materials because the Hole carriers dominate the electronic transport. High optical and thermal response for all compounds reveals that they are potential candidates for optical and thermoelectric devices.

Keywords: D-metal sulfides; P-type thermoelectrics; Optical analysis

---

Corresponding author email:

## 1. Introduction:

Semiconductors with the band gap in the energy range~1.5eV are perfectly considered as ideal materials for the solar cell development with high proficiency [1]. Moreover, the successful optoelectronic devices could be investigated with appropriate bad gap choice by focusing energy range. The appropriate energy band gap is the direct one where the immediate transition from valence to conduction band is possible for the productive optoelectronic applications. It is therefore necessary to find new materials with the ideal properties to fulfill the needs of the rapidly developing innovation.

The spinel materials $X_2YZ_4$, with both *X* and *Y* are (metallic elements) and *Z* belongs to (O, S, Se, Te) chalcogenides family, has motivated various researcher [2,3], since it exhibited significant physical properties i.e. phase transformations [4], half metallicity [5], colossal magnetoresistance [6], metal to insulator transitions [7], thermoelectricity [8], thermodynamic stability [9], charge storage ordering [10], transparent character for longer energy range [11], nonlinear optical susceptibility [12], etc., which make them promising candidates for assembling of optoelectronic devices and materials for various applications especially in magnetism, geophysics, catalysis and environment [13-16]. The thiospinels could be utilized for various applications in the field of defect engineering by implementing pressure in specific manners [17]

Peskov *et.al* reported the crystal chemical characterization of d-metal sulfides/selenides/tellurides containing discrete tetrahedral $TX_4$(X = S, Se, Te) groups with chalcogens anions and further occupied the similar structure as the binary compounds adopted [18]. The randomly scattered atoms in chalcogenides involve diverse crystallographic positions were not included initially in the sample. Thus, they got an information base containing structural information on 122 *d*-metal compounds with the formula $M_y[TX_4]$ (where M represent metal and y = 2 to 6). In these compounds the mechanisms behind the Inter-atomic bonds were majority resolved with the help of AutoCN program [19].

## 2. Computational details:

Different physical properties of the *d*-metal sulfides $Sc_2CdS_4$ and $Y_2CdS_4$ have been calculated using the state of the art density functional calculations. The full potential scheme of linearized augmented plane wave plus local orbitals is used as implemented in the Wien2K computer package [20]. The maximum angular momentum value $\boldsymbol{l}$ = 12 for the charge and potential expansion in spherical harmonics used. The radii of the muffin tin spheres are 2.21, 2.25, 2.50, and 2.15 bohr for Sc, Y, Cd, and S, respectively. For the interstitial region, the plane wave cut–off value ($K_{max}$ = 8.0/$R_{MT}$) was used and 1000 k-points is used for the Brilluion zone integrations. For the structural parameters determination of the materials, the generalized gradient approximation (GGA) [21] is used. On the other hand, the new semi-local potential known as the "Tran and Blaha modified Becke-Johnson" TB-mBJ [22] potential, used for the electric structure calculations due its better energy bandgap prediction compared to other less computational potentials.

## 3. Results and Discussion

### 3.1. Structural properties:

The ternary *d*-metal sulfides $Sc_2CdS_4$ and $Y_2CdS_4$ compounds exist in cubic structure with crystallographic parameters a=b=c, $\alpha = \beta = \gamma = 90^0$ and space group $Fd\bar{3}m$ (#227). The relaxed crystal structure (left) of the title compounds and $CdS_4$ tetrahedron (right) is shown in Fig.1. In this structure each Cd atom is tetrahedrally coordinated by four sulfur atoms, while Sc/Y atom is coordinated to six sulfur atoms. However, one sulfur atom shares by both Sc/Y and Cd atoms inside the unit cell, whereas there is no single bond present between Sc/Y and Cd atoms over the entire structure. The basic structure parameters of the compounds are determined by performing the volume optimization. The internal crystal structural parameters relaxation performed while considering the experimental lattice constant. Table1 depicts the calculated relaxed values of these parameters for both the compounds. A quick look at the table shows that the lattice constant increased by varying the monovalent cation, i.e., in the order Sc →Y; it is 10.85Å for $Sc_2CdS_4$ and 11.31Å for $Y_2CdS_4$. It is noticed that the lattice constant, unit cell volume and the bond length increases by replacing Sc with Y. This increase is due to the fact that Sc has larger atomic size than Y element.

**Table 1.** The calculated lattice constants *a, b* and *c* (in Å), bond angle α, β and γ, bond lengths *d* (in Å) and optimize volume (in $Å^3$), for $Sc_2CdS_4$ and $Y_2CdS_4$ compounds

| Compounds | *a =b= c* | *α =β= γ* | $V_o$ | *d* | *d* |
|---|---|---|---|---|---|
| **$Sc_2CdS_4$** | 10.85460 | 90 | 1278.914705 | 2.55195[Cd-S] | 2.60233[Sc-S] |
| **$Y_2CdS_4$** | 11.31190 | 90 | 1447.460389 | 2.59622[Cd-S] | 2.74566[Y-S] |

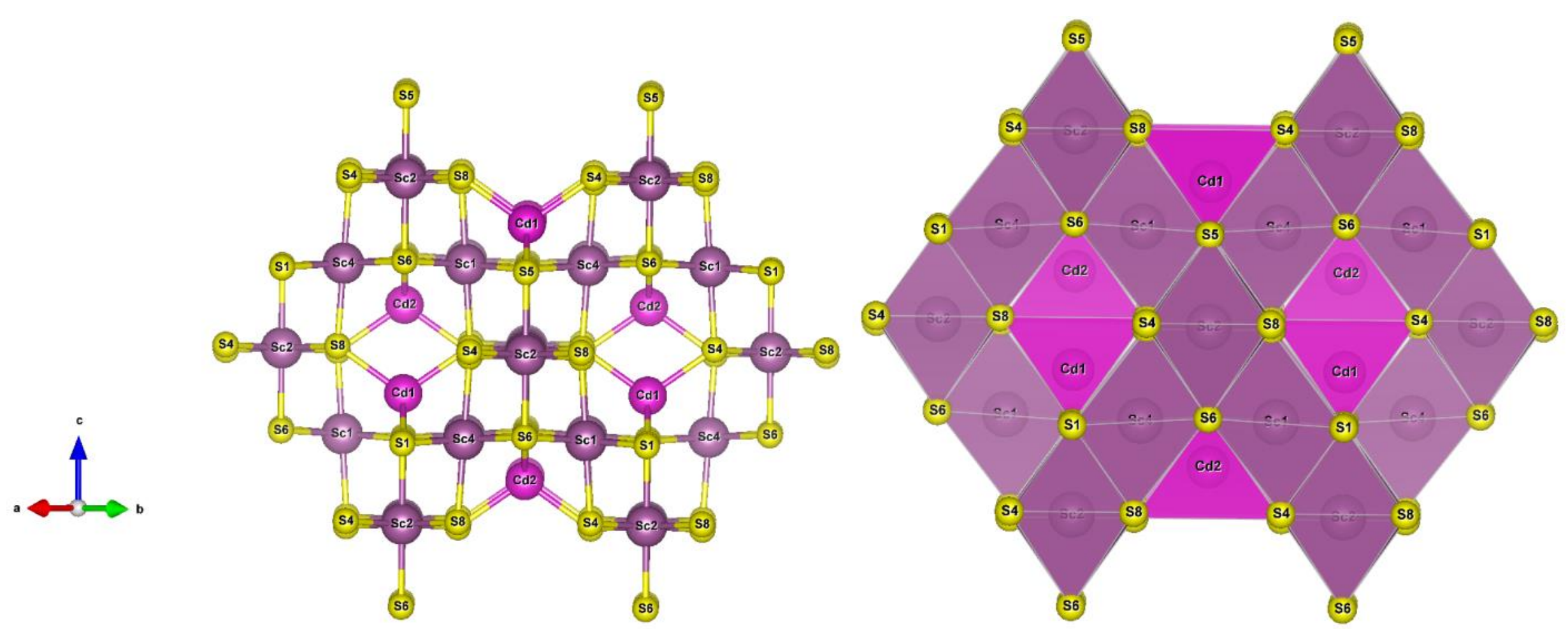

Fig.1. Relaxed crystal structure (left) and polyhedron (right) of $Sc_2CdS_4$ compound

### 3.2. Electronic Properties:

### 3.2.1. Energy Band Structure:

The calculated electronic band structures of $Sc_2CdS_4$ and $Y_2CdS_4$ compounds are displayed in Fig. 2, in the high symmetry direction W-L-Γ-X-W-K in the first Brillouin zone. The energy range is set from -4 to 4eV and the Fermi level is considered as reference (0eV). Both the compounds show semiconducting nature and direct band gap with a value 1.886eV for $Sc_2CdS_4$ and 2.209eV for $Y_2CdS_4$, respectively. The increase in the band gap is due to the replacement of lower atomic radius cation with the higher one. Due to the cation replacement bond length increased (shown in Table 1) which higher the energy band gap of the studied compounds. Despite no experimental value of the compounds is available in the literature, we believe that these values are more accurate and could be closer to the measured band gap of the materials. Therefore, from now on, only results obtained through Modified Becke-Johnson exchange (mBJ-GGA) potential are reported.

The effective mass of the compounds is also calculated from the bands structure at the Γ symmetry point. From the Table 2, it is clear that the electron effective masses are less than the hole effective masses. Further the $Y_2CdS_4$ has higher hole effective mass and lower electron effective mass comparatively to $Sc_2CdS_4$ due to its larger energy bandgap at the Γ point.

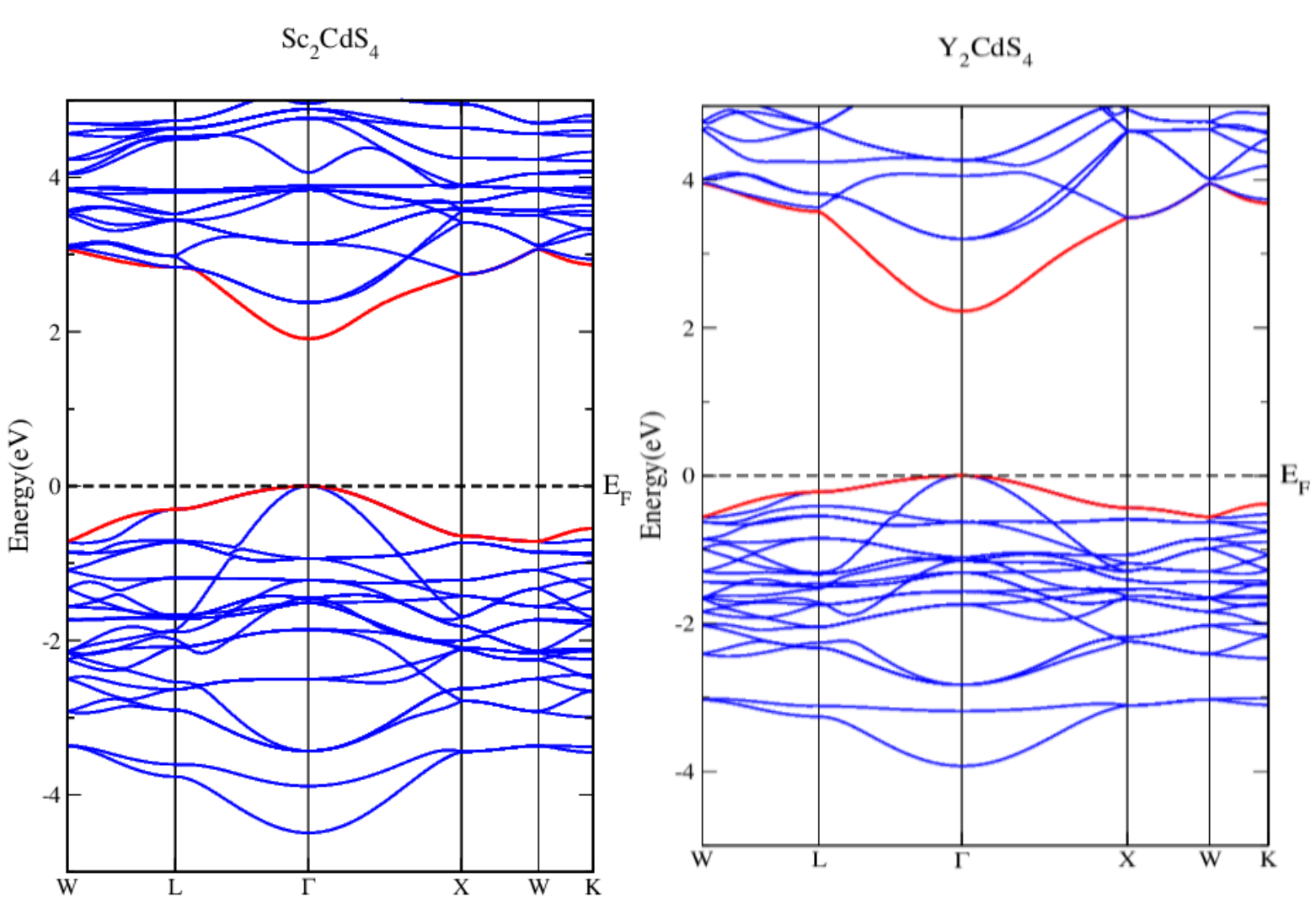

Fig. 2. Energy band structure calculated by mBJ-GGA of $Sc_2CdS_4$ and $Y_2CdS_4$ compounds.

**Table 1.** The calculated effective masses of the compounds (in units of a.m.u)

| Compounds | Electron effect mass ($m_e^*$) | Hole effective mass ($m_h^*$) |
|---|---|---|
| $Sc_2CdS_4$ | 0.0678*10^-31 | 0.2339*10^-31 |
| $Y_2CdS_4$ | 0.0575*10^-31 | 0.2823*10^-31 |

### 3.2.2. Density of states (DOS):

The total and partial densities of states (DOS) of $Sc_2CdS_4$ and $Y_2CdS_4$ compounds are illustrated in Fig. 3 in order to display the electronic states contribution of Sc/Y, Cd and S elements. From this figure we can say that the orbital contributions in both compounds are quite similar. For both these compounds close to the Fermi level in VBM (valence band region) towards higher energy the *p*-orbital of sulfur S atom shows predominant contribution while the *d*-orbitals of Sc/Y atom shows minor contribution. Moreover towards lower enegy in the valence band region near -7.24eV the sharp Cd-*d* and smaller S-*p* orbitals appear. Instead of this towards higher energy in CBM (conduction band region) starts near the band gaps of the compounds with the major contribution of Sc/Y-*d* and S-p,*d* orbitals with the lesser contribution of Cd-*d* orbitals in both compounds. Contribution of various states in conduction band region starts from the energy band gap of the under study compounds. The valence bands are remains on the Fermi level and conduction bands move away from the Fermi level in both these compounds as one move down in the periodic table from Sc to Y. From this investigation we may conclude that the direct energy bandgaps arise in the compounds $Sc_2CdS_4$ and $Y_2CdS_4$ are due to the strong hybridization between S-*p* and Sc/Y-*d* orbitals present among valence and conduction bands which is further beneicial to electrical transport. The major contrintion of S-p states in the entire valence band indicates the dominant ionic bonding character with small covalent part in these compounds.

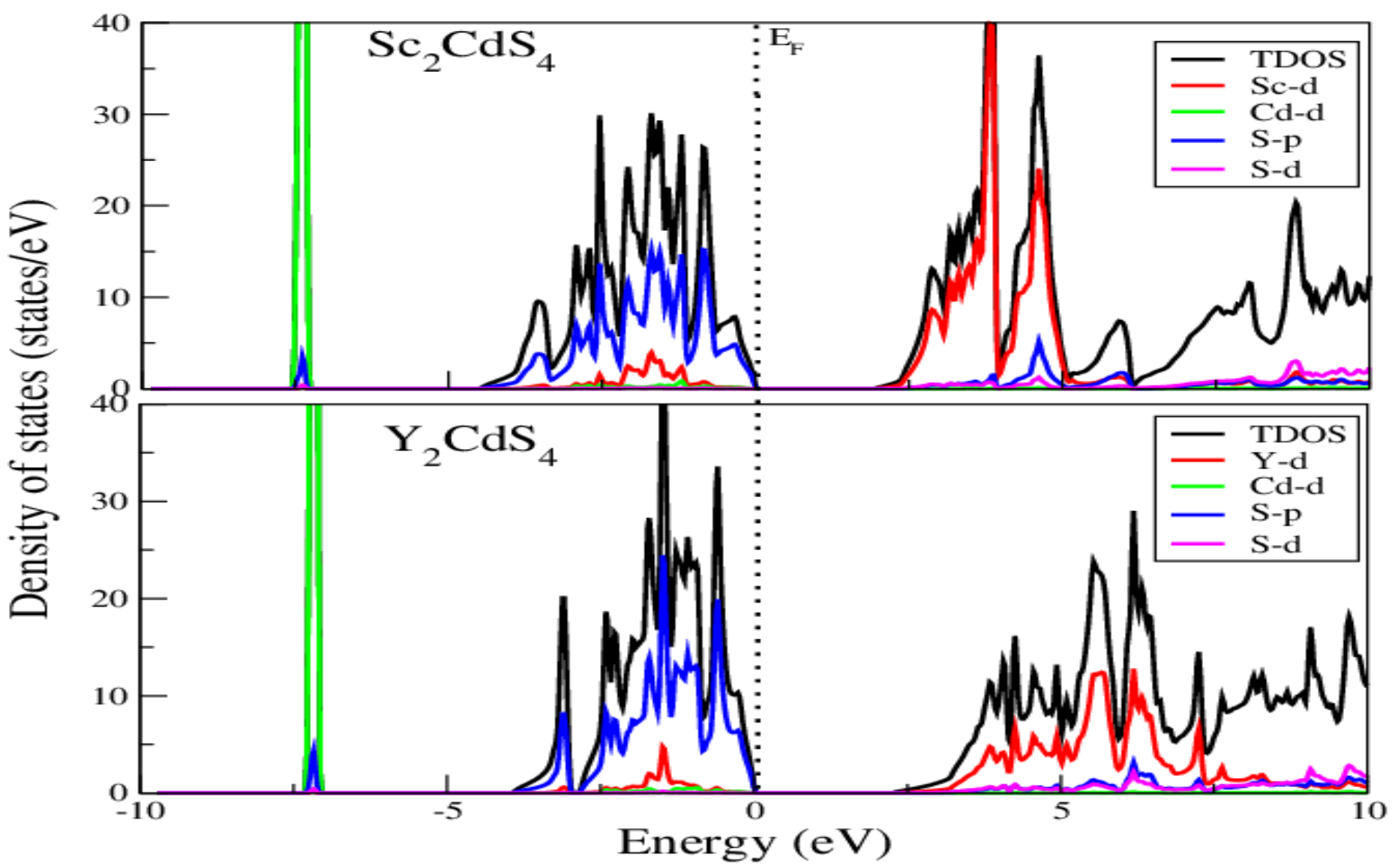

Fig. 3. Density of states (Total and Partial) of $Sc_2CdS_4$ and $Y_2CdS_4$ compounds calculated with mBJ.

### 3.3 Optical properties:

The linear optical responses of the compounds were reported in the energy range from 0-15eV as presented in Fig. 4(a-f). The optical response of a material at all photon energies can be expressed by dielectric function $\varepsilon(\omega) = \varepsilon_1(\omega)+i\varepsilon_2(\omega)$. The real part $\varepsilon_1(\omega)$ (denoting the polarizability of materials) and imaginary part $\varepsilon_2(\omega)$ comprised energy loss (absorption capability) of light photons by the materials for both the compounds. The dielectric function for $Sc_2CdS_4$ and $Y_2CdS_4$ compounds were calculated in energy range of 0-15eV and are presented in Fig. 4a. The spectrum of $\varepsilon_1(\omega)$ for $Sc_2CdS_4$ and $Y_2CdS_4$ compounds start at low frequency with value 5.98 and 5.01, which is termed as static dielectric function. It is clearly understand from the figure, that band gap and the static dielectric constant $\varepsilon_1(0)$ of the compounds are strongly related. This trend is clearly observing in a compounds energy band gap values, that the smaller band gap produced a larger $\varepsilon_1(0)$ value, when one moves down in the group from Sc to Y followed the Pen model. After low frequency values, $\varepsilon_1(\omega)$ increase to maximum value of 11.13 at 3.32eV energy for $Sc_2CdS_4$ and 9.14 at 3.97eV energy for $Y_2CdS_4$, respectively. The main characteristic peaks of the under study compounds located in the visible energy range of electromagnetic spectrum. Beyond maximum peaks, $\varepsilon_1$ decreases sharply below zero beyond the energy 5.68 and 7.10eV for $Sc_2CdS_4$ and $Y_2CdS_4$ compounds. The compounds exhibited greater energy losses beyond these values after certain energy ranges, depicts complete attenuation of light in the optical medium and show metallic like behavior of the material. For negative values of $\varepsilon_1$ the material shows high reflectivity. Moreover, we observed that the peaks moved toward higher energies due to the increase in the band gap when Sc cation replace by Y.

The imaginary part $\varepsilon_2(\omega)$ is sum of all transition from the valence to conduction bands. Thus it gives information about absorption behavior of $Sc_2CdS_4$ and $Y_2CdS_4$ compounds as shown in Fig. 4a. The curves for $\varepsilon_2(\omega)$ show that its starting point is $E_o$= 2.10eV for $Sc_2CdS_4$ and 2.34 for $Y_2CdS_4$. As electrons transition take place between different band (inter band transition) due to which several peaks are detected further than threshold point. It can be characterized by transition of electrons from Cd-*d* and S-*p* states in the valence band to Sc/Y-*d* states in the conduction band for $Sc_2CdS_4$ and $Y_2CdS_4$, respectively. After the threshold points, the curve increases monotonically due to the enhanced joint density of states contributing to $\varepsilon_2(\omega)$. The dominant two peaks appear in both these compound. The maximum absorption capabilities with the main peak of $Sc_2CdS_4$ and $Y_2CdS_4$ have magnitude 9.704 and 8.734 at energy 5.315 and 6.820eV, respectively. The reported compounds peak lies in the ultraviolet region while it shifted towards higher energy and become low in going from Sc to Y. Being a medium band gap material $Sc_2CdS_4$ and $Y_2CdS_4$ compounds, showed large absorption covering the entire far visible region together with the near and up to some extent far ultraviolet photons. Therefore, the stable and UV-Vis absorbent are promising candidates for solar cell technology.

The calculated plot for refractive index $n(\omega)$ is displayed in Fig. 4b. The solid lines represent $Sc_2CdS_4$ while the broken lines represent $Y_2CdS_4$. Interestingly, $n(\omega)$ follows the same behavior as $\varepsilon_1(\omega)$. This similar trend of both spectra is supported with the established theory. The static value $n(0)$ of refractive index was found to be 2.44 for $Sc_2CdS_4$ and 2.24 for $Y_2CdS_4$ follows the same behavior as $\varepsilon_1(0)$. The highest peaks of refractive index $n(\omega)_{max}$ are appeared with a value 3.407, 3.096 at energy 3.36, 4.04eV for $Sc_2CdS_4$ and $Y_2CdS_4$ compounds respectively, which are in visible region of electromagnetic spectrum. It is clearly observed that the small band gap compounds have high refractive index. These results show that $n(0)$ and $n(\omega)_{max}$ decreases by replacing Sc with Y which means it follows the same trend as $\varepsilon_1(0)$ while opposes the trend present in energy band gap of the compounds. After attaining the maximum peak values with the minor peak appears on right side of the spectra of both the compounds it starts decreasing because after getting the peak value the material cannot absorb more incident light photons. This dominant decrease in the peak indicates dispersion with photon energy. Clearly for these compounds, refractive indices are large in the visible region.

Extinction coefficients $K(\omega)$ tell us about the absorption capability of the material when incident light pass through it. $K(\omega)$ depends upon the energy of incident photons and nature of the material. The extinction coefficient $K(\omega)$ also follows the same pattern as that of $\varepsilon_2(\omega)$ as shown in Fig. 4b. The value of $K(\omega)$ starts at the fundamental band gaps of the compounds then abruptly increases and reach at its maximum peak values with the minor peak appears on left side of the spectra of both the compounds. The maximum reported values of extinction Coefficients $K(\omega)max$ are 2.044, 1.941 at energy 5.46, 6.87eV for $Sc_2CdS_4$ and $Y_2CdS_4$ compounds respectively. The $Sc_2CdS_4$ show the maximum value of extinction coefficient which means it absorbs more light as compared to $Y_2CdS_4$ compound. Moreover, the maximum peaks move towards the higher energies whereas a clear multi peaks with shoulders appear in both compounds as well by replacement of the cations from Sc to Y. This is due to the increase in the energy band gap of the compounds in going from Sc to Y.

The frequency-dependent reflectivity $R(\omega)$ results are shown in Fig. 4c. The static frequency limit of reflectivity $R(0)$ for $Sc_2CdS_4$ is 17% and $Y_2CdS_4$ is 14% respectively follows the same behavior as $\varepsilon_1(0)$, and these compounds exhibited maximum reflectivity values of 39% ($Sc_2CdS_4$) and 38% ($Y_2CdS_4$) in the energy range from 2.5–15eV. Beyond maximum peaks, $\varepsilon_1(\omega)$ goes below zero to revealing the metallic character beyond the energy 5.68 and 7.10eV for $Sc_2CdS_4$ and $Y_2CdS_4$ compounds. The maximum calculated values of reflectivity $R(\omega)$ lie in ultraviolet energy range. These results reveal that the materials can be used as shielding materials in the energy range from 2.5–15eV due to their high achieving value of reflectivity in this region. Further, it is clear from the spectra of $R(\omega)$ that the static values decrease by changing the cation from Sc to Y. It is due to the fact that the band gaps of compounds are increased, which also broadens the high reflectivity range of the compounds. From the plot we can see that maximum reflection occurs by $Sc_2CdS_4$ compound.

The optical conductivity $\sigma(\omega)$ is computed to obtain the electronic characteristics for $Sc_2CdS_4$ and $Y_2CdS_4$ compounds as shown in Fig. 4d. The optical conductivity $\sigma(\omega)$ starts at the band gap value of the under study compounds. Further increase in energy causes considerable increase in the $\sigma(\omega)$ value. The $\sigma(\omega)$ increases slowly with smaller peaks for both the understudy compounds and then reaches gradually to maximum peak value, the extreme maximum values $\sigma(\omega)_{max}$ are 6985.19, 7845.69 $\Omega^{-1}cm^{-1}$ for $Sc_2CdS_4$ and $Y_2CdS_4$ compounds at energy 5.38, 6.87eV, respectively. As from the plot several peaks are observed in the energy range between 3eV up to 12.5eV which predicts that optical conductivity occurs at different points. $Y_2CdS_4$ compound possess greatest $\sigma(\omega)$ than $Sc_2CdS_4$. After reaching maximum values it starts decreasing with multiple minor peaks in the high energy range. Compounds show prominent optical conductivity in the UV region, consequently, these systems can be functional for the high frequency devices operating in this range.

The absorption coefficient $\alpha(\omega)$ is important in describing the intensity of the light in the medium and optical density. Fig. 4e depicts the absorption coefficient spectra of $Sc_2CdS_4$ and $Y_2CdS_4$ compounds. It starts at the critical point and then increases above these compounds with characteristics optical structures originated due to the electronic transitions with high joint density of states. The maximum values of $\alpha(\omega)max$ for $Sc_2CdS_4$ and $Y_2CdS_4$ compounds are 175.71, 148.52×$cm^{-1}$ at energy 12.1, 13eV, respectively. $Sc_2CdS_4$ compound possess greatest $\alpha(\omega)$ than $Y_2CdS_4$. It shows that the absorption strength is decreasing by replacing Sc with Y for the compounds.

The sum rule is important parameter which estimates effective number of electrons per unit cell in the valence band that take part in both inter and intra-band transition. It is obvious from Fig. 4f that electron take contribute in transitions at round 3.50eV and 4.07eV for $Sc_2CdS_4$ and $Y_2CdS_4$ and having no appearance up to the energy band gap of the compounds. Beyond these values, $N_{eff}$ increases for higher energies, further this highly affects the optical parameters.

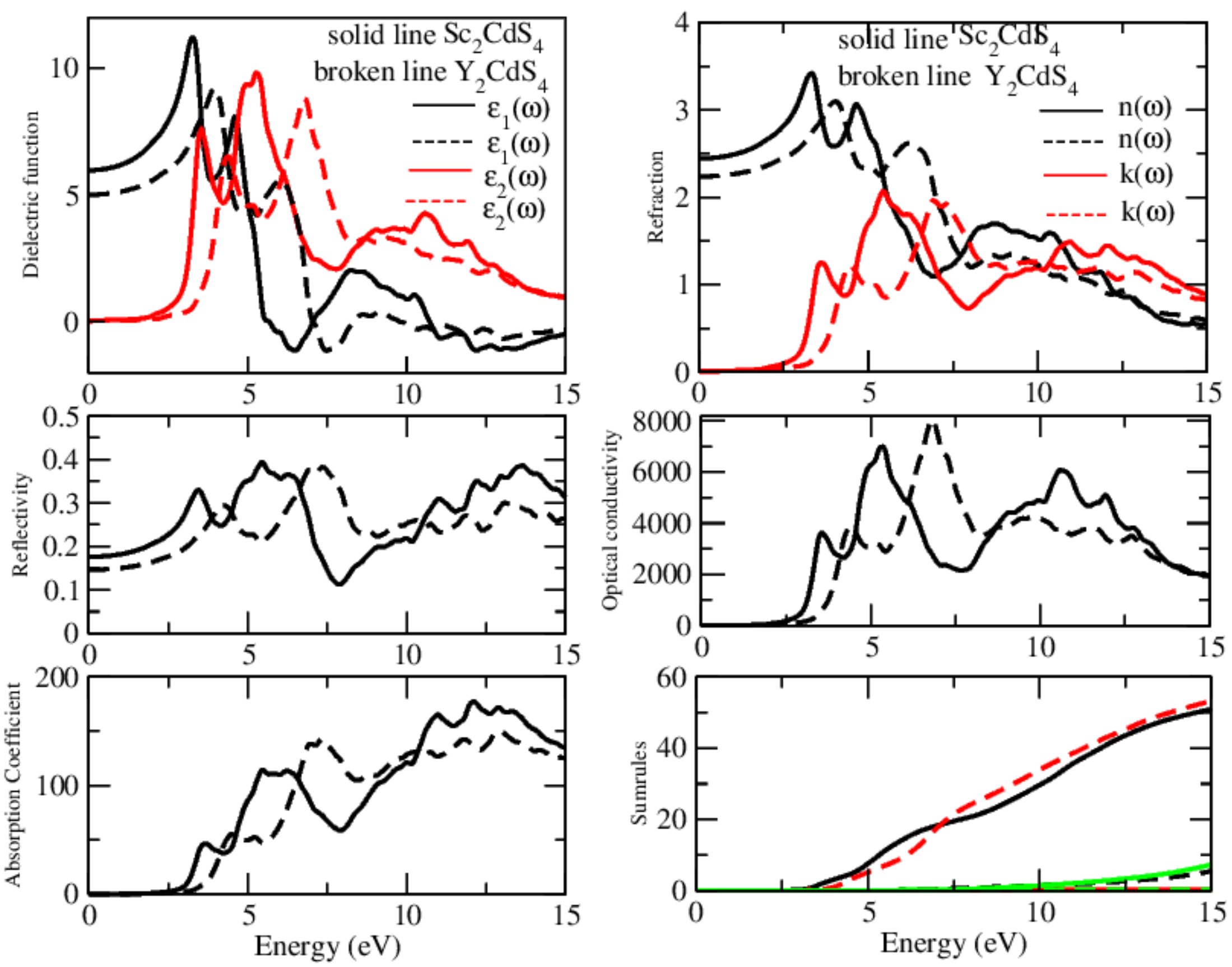

Figure 4 (a-f). Optical parameters of $Sc_2CdS_4$ and $Y_2CdS_4$ compounds in the energy range between 0 to 15eV. The solid lines represent $Sc_2CdS_4$ while the broken lines represent $Y_2CdS_4$.